\documentclass[5p]{elsarticle}
\usepackage{lineno,hyperref}
\modulolinenumbers[5]
\usepackage{graphicx}
\usepackage{amsmath}
\usepackage{natbib}
\usepackage{url} 
\usepackage{dblfloatfix}  
\modulolinenumbers

\journal{Journal of \LaTeX\ Templates}
\usepackage{color}

\usepackage{numcompress}\biboptions{authoryear}

\begin{document}

\begin{frontmatter}

\title{SOME ADDITIONAL RESULTS OF THE HILDA GROUP ORBITAL ELEMENTS APPROXIMATIONS}

\author[1]{Rosaev, A.}
\ead{hegem@mail.ru}


\cortext[cor1]{Corresponding author}

\address[1]{Research and Educational Center "Nonlinear Dynamics", Yaroslavl State University, Yaroslavl, Russia}
\begin{abstract}

Context: The quantitative characterization of amplitudes and periods of perturbations in orbital elements can help to our understanding the minor bodies dynamics, especially in case motion in vicinity of resonances. 

Aims: The main goal of this study is to present the new way of the orbital elements approximation by the results of numerical integration.

Method: We develop the approximation of orbital elements of the small body by the perturbations with combinational frequencies. We use a criterion of minimum of the standard error in our approximation. 

Result: In result we give the approximation of orbital elements for the selected members of the Hilda group in the 3:2 mean motion resonance with Jupiter. 
 
\end{abstract}
\end{frontmatter}


\section{Introduction}

The solution for the n-body problem has long eluded us. It is commonly assumed, that the n-body problem is not integrable in general, even in simplified settings (\cite{MD}, \cite{Me1}), i.e. it is impossible to obtain a solution in the form of analytical functions or their combinations. Almost all previous attempts centred around a solution employing differential equations of motion or their simplifications. However, another approach can be examined: to approximate the results of numerical integrations by a quasi-periodic function, considering the stability of this solution and then generalizing it to a (small) finite volume. It would then be possible to establish a relation with a defined perturbation affecting the system (or particle). 

{The problem is still relevant: existing theoretical models give different results depending on the intended simplifications (see for example \cite{BM}).if successful, the quantitative approximation of the results of numerical integration can help to choice between theoretical models.}
 
{Consider the classical problem of n-bodies with some limitations: one mass is much larger than the other, the orbits of the planets do not intersected. This can be called the planetary n-body problem.}

Our ultimate goal is to perform the motion of an infinitesimal mass (asteroid) in the gravitational field of the central mass (star) and several planets in orbits close to circular ones with maximum accuracy.

First of all, it is necessary to consider the known approximation methods.  
The one of example is well known: harmonic (Fourier) approximation (\cite{La}). However, it is valid only in a finite interval and significantly depends on the (arbitrary) choice of the integration interval. 


In our previous paper \cite{RP5}, we applied another type of approximation which calculated multiple perturbations with different (incommensurable) frequencies. Here we propose a new type of approximation applicable in the case of resonance.

The Hilda group is a unique array of orbits in the exact (3:2) mean motion resonance which is stable for a long time. In other resonances in the main belt, we have gaps that asteroids rarely visit. The origin of this group is connected with the process of planetary migrations \cite{NV2}. For these reasons, a detailed study of the dynamics of this group is very important. The present paper provides some new quantitative data and draws attention to the difference in the dynamical state of the asteroids within the Hilda group. 


\begin{table*}[t]
\caption{The synthetic proper elements for the selected Hilda group members. Data of elements 10.02.2021}
\begin{center}
\begin{tabular}{l l r r r r r r r r} 
\hline 
 \multicolumn{2}{| c |}{Asteroid }  &   \multicolumn{5}{c |}{Proper elements}  &   \multicolumn{2}{c |}{Present paper}   \\
 \multicolumn{2}{| c |}{}&    \multicolumn{1}{c |}{a [au]}   &  \multicolumn{1}{c |}{e}  &  \multicolumn{1}{c |}{$LCE$ }  &  \multicolumn{1}{c |}{$s,''/yr$}  &  \multicolumn{1}{c |}{$g,''/yr$}  &  \multicolumn{1}{c |}{$s,''/yr$}  &  \multicolumn{1}{c |}{$g,''/yr$}  \\
  \hline \hline 
\multicolumn{1}{| c }{(153)} & \multicolumn{1}{l |}{Hilda}   & \multicolumn{1}{c |}{3.96525}  & \multicolumn{1}{c |} {0.155005}  & \multicolumn{1}{c |} {1.27}   &     \multicolumn{1}{c |}{-63.1393}    &     \multicolumn{1}{c |}{-490.585}  &  \multicolumn{1}{c |}{-63.00}  &  \multicolumn{1}{c |}{-489.60} \\
\multicolumn{1}{| c }{1180} & \multicolumn{1}{l |}{Rita}  & \multicolumn{1}{c |}{3.96485}  & \multicolumn{1}{c |} {0.10729}  & \multicolumn{1}{c |} {1.42}   &     \multicolumn{1}{c |}{-72.5112}    &     \multicolumn{1}{c |}{-535.231} &  \multicolumn{1}{c |}{-72.36}  &  \multicolumn{1}{c |}{-534.46} \\
\multicolumn{1}{| c }{3990} & \multicolumn{1}{l |}{Heimdal}  & \multicolumn{1}{c |}{3.96507}  & \multicolumn{1}{c |} {0.168435}  & \multicolumn{1}{c |} {1.40}   &     \multicolumn{1}{c |}{-67.6059}    &     \multicolumn{1}{c |}{-513.665} &  \multicolumn{1}{c |}{-67.57}  &  \multicolumn{1}{c |}{-513.32} \\
\multicolumn{1}{| c }{248430} & \multicolumn{1}{l |}{2005 SM284}  & \multicolumn{1}{c |}{3.96628}  & \multicolumn{1}{c |} {0.195248}  & \multicolumn{1}{c |} {2.79}   &     \multicolumn{1}{c |}{-56.4352}    &     \multicolumn{1}{c |}{-382.992} &  \multicolumn{1}{c |}{-56.52}  &  \multicolumn{1}{c |}{-382.93} \\
\multicolumn{1}{| c }{6124} & \multicolumn{1}{l |}{Mecklenburg}  & \multicolumn{1}{c |}{3.96613}  & \multicolumn{1}{c |} {0.191510}  & \multicolumn{1}{c |} {5.69}   &     \multicolumn{1}{c |}{-59.6946}    &     \multicolumn{1}{c |}{-400.028} &  \multicolumn{1}{c |}{-59.40}  &  \multicolumn{1}{c |}{-399.38} \\
\multicolumn{1}{| c }{1911} & \multicolumn{1}{l |}{Schubart }  & \multicolumn{1}{c |}{3.96586}  & \multicolumn{1}{c |} {0.190597}  & \multicolumn{1}{c |} {1.61}   &     \multicolumn{1}{c |}{-62.0142}    &     \multicolumn{1}{c |}{-423.648} &  \multicolumn{1}{c |}{-61.56}  &  \multicolumn{1}{c |}{-423.50} \\
\multicolumn{1}{| c }{448203} & \multicolumn{1}{l |}{2008 UH167}  & \multicolumn{1}{c |}{3.96555}  & \multicolumn{1}{c |} {0.183724}  & \multicolumn{1}{c |} {0}   &     \multicolumn{1}{c |}{-66.2933}    &     \multicolumn{1}{c |}{-457.903} &  \multicolumn{1}{c |}{-66.24}  &  \multicolumn{1}{c |}{-457.88} \\
\multicolumn{1}{| c }{46629} & \multicolumn{1}{l |}{1994 PS38}  & \multicolumn{1}{c |}{3.96552}  & \multicolumn{1}{c |} {0.183341}  & \multicolumn{1}{c |} {1.46}   &     \multicolumn{1}{c |}{-65.755}    &     \multicolumn{1}{c |}{-460.273} &  \multicolumn{1}{c |}{-65.59}  &  \multicolumn{1}{c |}{-460.55} \\
\multicolumn{1}{| c }{958} & \multicolumn{1}{l |}{Asplinda}  & \multicolumn{1}{c |}{3.96516}  & \multicolumn{1}{c |} {0.172800}  & \multicolumn{1}{c |} {2.04}   &     \multicolumn{1}{c |}{-73.1892}    &     \multicolumn{1}{c |}{-504.699} &  \multicolumn{1}{c |}{-73.22}  &  \multicolumn{1}{c |}{-505.08} \\
 \hline 
\end{tabular}
\end{center}
\label{tabH32Proper}
\end{table*}
We are testing four asteroids of the Hilda subfamily, three asteroids of the Schubart subfamily and two asteroids(1180 Rita and 958 Asplinda) with intermediate orbital elements. There are two close pairs of orbits: 6124 Mecklenburg and 248430(2005 SM284) in the Hilda subfamily and 46629 (1994 PS38) and 448203 (2008UH167) in the Schubart subfamily. (Table \ref{tabH32Proper}, where the proper elements according \cite{KM} are compared with mean elements by our integration). 

Throughout the article we have used standard notations for orbital elements, a -- semi-major axis in a.u., e~--~eccentricity, i -- inclination, $\Omega$ -- longitude of ascending node, $\varpi$ -- argument of perihelion(the angular elements are in degrees). 

\section{Method}

We start with our method which is described in details previously \cite{RP5}. Here we repeat the main states of it. The evolution of the orbital elements of asteroids is obtained by numerical integration. 
To study the long term dynamical evolution of orbits, the equations of the motions of the systems were numerically integrated backwards 800 kyrs, using the N-body integrator Mercury \cite{Ch1} {with} the Everhart integration method \cite{E1}. To avoid short periodic perturbations, we use a running-box averaging the results of numerical integration in a 500 year window.  The initial epoch of our integrations was $T_0=$ 1998 July 06 ($JD2451000.5$).

As it is well known, there is a possible approximation of the evolution of the orbital element (E) by a finite sum given by the partial Fourier series with some defined precision:
\begin{align}
E_i=E_{i0}t+\frac{c_{i0}}{2}+\sum_{k=1}^{N} c_{ik} \cos\left( k\omega_{ik} t + \phi_{ik} \right) ,
\label{Ei}
\end{align}
where $c_{i0}$/2 is the coefficient of the constant function (if $k=0$ then $\cos k \omega_it$=1).  Secondly, we use the approximation by a sum of trigonometric terms with arbitrary frequencies:
\begin{align}
E_i=&E_{i00}+E_{i0}t+\sum_{k=1}^{N}  c_{ik} \cos \left( f_{ik}t+ \Phi_{ik} \right) \ .
\label{Ei_new_new}
\end{align}
Here coefficients $c_{ik}$ -- amplitude, $f_{ik}$ -- frequency and $\Phi_{ik}$ -- phase. 
{$E_{i0}$ always equals zero for eccentricity, inclination and semi-major axis.}

The final approximation reached when the standard error which calculated using the expression:  

\begin{equation}
\sigma=\sqrt{\frac{1}{n-2}\sum_{i}^{n}\left( E_i-E_{iApprox} \right)^2}
\label{sig}
\end{equation}

when it becomes minimal.

The first possible application of our approximation is an independent (quantitative) estimation of the age of close asteroid pair and families. An example of this application is considered in our papers \cite{RP}. The second important application is the study different resonance perturbations in the evolution of orbital elements of asteroids.  The first attempt of such study was done in paper \cite{R2}.  Here we describe the other form of approximation, applicable in resonance case. 

However, not all resonance perturbations can be approximated by the expression (\ref{Ei_new_new}). For example, evolution of the Europa, one of Galilean satellite of Jupiter, displayed modulated oscillations \cite{Ce1}.  Below we listed some more examples of similar behavior in the Hilda group of asteroids in the 3:2 resonance with Jupiter. 

Obviously, the more complex approximation is required. For the better fit such complex cases it is natural to apply the approximation by the combinational frequencies:

\begin{align}
E_i=&E_{i00}+E_{i0}t+ d_{i} \cos \left( f_{i}t+ \Phi_{i} \right) \cos \left( \nu_{i}t+ \Psi_{i} \right) .
\label{Ei_mult}
\end{align}

As above, the final approximation reached when the standard error which calculated using the expression (\ref{sig}) obtained minimal value. 
For the target of the approximation, the original program was written. Obviously, it is possible to combine the approximations (\ref{Ei_new_new}) and (\ref{Ei_mult}).

\begin{align}
\nonumber
E_i=E_{i00}+E_{i0}t+\sum_{k=1}^{N}  c_{ik} \cos \left( f_{ik}t+ \phi_{ik} \right)+
\\
d_{i} \cos \left( f_{i}t+ \Phi_{i} \right)\cos \left( \nu_{i}t+ \Psi_{i} \right) .
\label{Ei_multi}
\end{align}


To better characterization of perturbations in range up to 1 Myr the approximations in two intervals 100 kyr and 800 kyr is applied. The results are given below in Tables (\ref{tabinc})-(\ref{tabPeri}).



\section{Some results of orbital elements approximations}

\subsection{Result of approximation in 800-kyr interval}
First, we have made the approximation of 153 Hilda orbital elements by method \cite{RP5} in the 800 kyr interval. The method is successful for inclination and node longitude (t in kyr):
  
  \begin{align}
  \nonumber
  \Omega =-231.4^o-17.75 t+9.79\cos(0.3071t+6.13)+
  \\
  +2.13\cos(0.0057t+5.85)
  \label{Omega-appr}
  \end{align}
  The standard error after the first (linear approximation is $\sigma_1=6.847$, after the second approximation $\sigma_2=2.646$  and after the third ones $\sigma_3=2.587$. The value 17.55 deg/kyr is equal to 63 arcsec/yr (in according with data in Table \ref{tabH32Proper}).
    \begin{align}
    \nonumber
    i =9.01^o-1.5595\cos(0.30695t+4.3809)+
    \\
    +0.475\cos(0.1786t+4.48)
    \label{i-appr}
    \end{align}

\begin{figure}
               \centering 
                 \includegraphics[width=8.7cm]{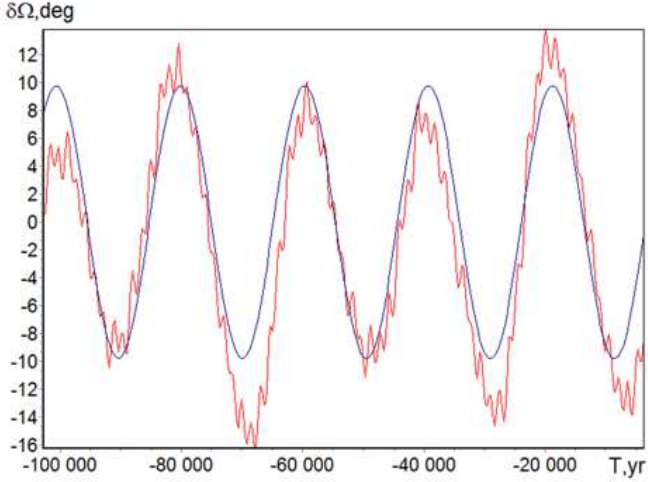}
                 \caption{The node longitude evolution of 153 Hilda after the subtraction of the mean orbital plane rotation}
            \label{Hnode}
             \end{figure}

 As an example, we give the approximation of node longitude of  the 153 Hilda variations (Figure \ref{Hnode}).

\subsection{Result of approximation in 100-kyr interval}

\begin{figure}
          \centering 
            \includegraphics[width=8.7cm]{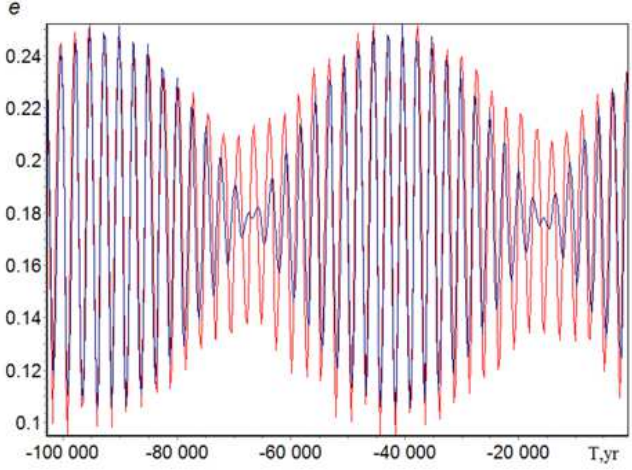}
            \caption{The eccentricity evolution of 153 Hilda}
       \label{Hex}
        \end{figure}

\begin{figure}
                \centering 
                  \includegraphics[width=8.7cm]{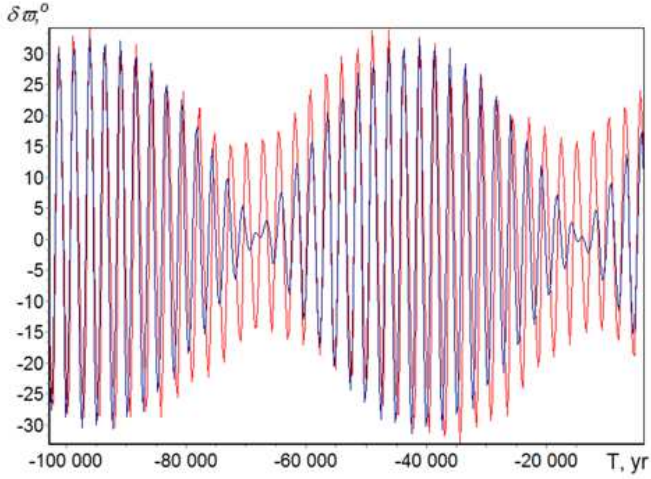}
                  \caption{The perihelion longitude evolution of 153 Hilda after the subtraction of the mean perihelion rotation}
             \label{Hw}
              \end{figure}

  The evolution of eccentricity and perihelion longitude is complicated to modulated perturbations. By this reason we apply other averaging interval for these variables (100 kyr) and new type of approximations, described above (expression(\ref{Ei_mult})).
  We have obtained (see Figures \ref{Hex} and \ref{Hw}):
  
  \begin{align}
      \nonumber
      e =0.18+0.074\cos(0.0615t+5.675)*
      \\
      *\cos(2.46t+1.19)
      \label{e-appr}
      \end{align}

 The standard error after this approximation is $\sigma_1=0.0177$,
  
  \begin{align}
      \nonumber
      \varpi =245.0^o-136.35t+25\cos(0.0585t+5.55)*
      \\
      *\cos(2.46t+4.0)
      \label{w-appr}
      \end{align}

 The standard error after the first (linear) approximation is $\sigma_1=15.6$, after the second approximation $\sigma_2=10.2$. The value 136.35 deg/kyr is equal to 490.9 arcsec/yr (in agreement with data in AstDys site, see Table \ref{tabH32Proper}).
 
 As it is seems, one of the perturbation periods in eccentricity is equal to ones in perihelion longitude. Similarly, the periods in inclinations and node longitude are the same.
 
  \cite{BF} obtain the same dependence for the periods of the perihelion longitude for case two exoplanets close to resonance (see Figure 6 in cited paper). Possible, this dependence is very common. However, it is necessary to note that studied perturbation in the perihelion longitude is less clear than in eccentricity.
 
 The same type of approximation is true for other members of Hilda group (Tables (\ref{tabinc})-(\ref{tabPeri})). The accuracy of the approximation of the coefficients in Tables (\ref{tabinc})-(\ref{tabPeri}) is bounded by the last decimal digit.  The value of sigma characterizes the residuals and may be considered as the precision of our fit on average for each element. The mean value of the smallest frequency of the perturbation in eccentricity is about $0.059\pm0.001 kyr^{-1}$; in the same time the higher frequency is varying in the larger range about 30$\%$.
 
 The Hilda and Schubart families are clearly separated not only by the mean inclination values but also by the amplitudes of the node longitude perturbation. The amplitudes of the node longitude perturbation in case of the Schubart family are about three times larger. But noted difference concerns not only the Hilda and Schubart families. The coefficients, obtained during the approximation (i.e. periods, amplitudes and phases) are slightly different for different members of the Hilda group (see Tables (\ref{tabinc})-(\ref{tabPeri})). 
 
 The reason for this is currently unknown. Perhaps this may be due to the features of the origin of the studied asteroids on their current orbits. We believe that this is important and required a future theoretical explanation. The coefficients, obtained for two close pairs can help to understand the problem of (possible) their common origin and the possibility of fragmentation in the Hilda group. However solving this problem requires taking into account some other factors such as the Yarkovsky effect and is out of target of this paper.
 {According to \cite{BV}, the effect of thermal force in case 3:2 resonance may be differ than in the non-resonance part of the Main Belt.}
 
 To verify our results, we repeated our approximation in larger interval (800 kyr). In general, this test confirmed the values of periods and amplitudes of perturbations, determined in the 100 kyr interval (compare expression above (\ref{i-appr}) and (\ref{Omega-appr}) with data in Tables (\ref{tabinc}) and (\ref{tabNod})). 
 
 
 \subsection{Result of Fourier approximation}
 
 To comparison, we applied the approximation of the orbital elements evolution for the 153 Hilda by Fourier frequencies.
 
 First, we use the 100 kyr interval for the Fourier approximation. In result we obtain the frequencies of main perturbations in eccentricity $2.39\pm 0.03 kyr^{-1}$  and $2.51\pm 0.03 kyr^{-1}$. 
 After that we use the 800 kyr interval for the Fourier approximation. It gives the better result. Finally, we obtain the frequencies of main perturbations $2.392\pm 0.005 kyr^{-1}$ and $2.510\pm 0.004 kyr^{-1}$. (the ratio between amplitudes of two main harmonics is 2.73). The distribution of the amplitudes in the Fourier spectra can tell about strong nonlinearity of perturbations.

 By our method we obtain the values $2.399\pm 0.001 kyr^{-1}$ and $2.522\pm 0.001 kyr^{-1}$ respectively. Our values are in range of errors obtained by Fourier method but more precise.  Similar conclusion is true for the phase of perturbations. Additionally, two perturbations with large amplitude are detected by Fourier method are obviously the second harmonics. However, the distribution of the amplitudes in the Fourier spectra (ratio between amplitudes of two main harmonics is 3.25) is important and notes to the nonlinear character of  the perturbation. 
 
\begin{figure}
      \centering 
       \includegraphics[width=8.7cm]{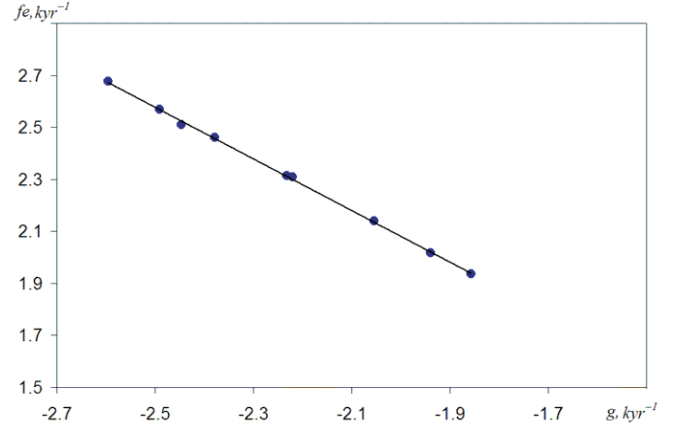}
        \caption{The dependence of eccentricity main frequency on proper frequency}
   \label{FeFg}
   
    \end{figure}

\begin{figure}
       \centering 
        \includegraphics[width=8.7cm]{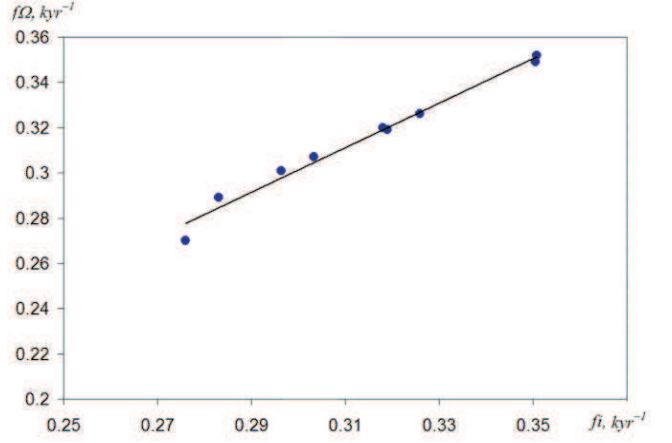}
         \caption{The dependence of node longitude main frequency on inclination frequency}
    \label{FeFs}
     \end{figure}

  {The higher frequency, detected in eccentricity evolution is the same as proper frequency $g$ (Figure \ref{FeFg}). Similarly, the main frequency detected in inclination and node longitude coincide with proper frequency $s$ (Figure \ref{FeFs}) . Previously we obtain the same results for case non-resonance motion.} 
   
 \begin{table*}[t]
 \caption{The inclination evolution approximation for the selected Hilda group members. Integration interval 100 kyr}
 \begin{center}
 \begin{tabular}{l l r r r } 
 \hline 
  \multicolumn{2}{| c |}{Asteroid }  &   \multicolumn{2}{c |}{}  \\
  \multicolumn{2}{| c |}{}&    \multicolumn{1}{c |}{$i$}    &  \multicolumn{1}{c |} {$\sigma $}  \\
   \hline \hline 
 \multicolumn{1}{| c }{} & \multicolumn{1}{l |}{153 Hilda}   & \multicolumn{1}{c |}{$ 9.03+  1.55\cos( 0.3033t+  3.95)$}  & \multicolumn{1}{c |} { 0.32962}   \\
 \multicolumn{1}{| c }{} & \multicolumn{1}{l |}{1180  Rita} & \multicolumn{1}{c |} {$ 6.26+  1.47\cos( 0.3506t+  0.27)$
 }  & \multicolumn{1}{c |} { 0.32173}   \\
 \multicolumn{1}{| c }{} & \multicolumn{1}{l |}{3990 Heimdal} & \multicolumn{1}{c |} {$ 9.65+  1.53\cos( 0.3259t+  4.52)$
 }  & \multicolumn{1}{c |} {0.33444}  \\
 \multicolumn{1}{| c }{} & \multicolumn{1}{l |}{248430(2005 SM284)} & \multicolumn{1}{c |} {$  9.04+  1.40\cos( 0.2760t+  0.72)$
 }  & \multicolumn{1}{c |} { 0.38762}    \\
 \multicolumn{1}{| c }{} & \multicolumn{1}{l |}{6124 Mecklenburg} & \multicolumn{1}{c |} {$  8.83+  1.60\cos( 0.2830t+  5.00)$
 }  & \multicolumn{1}{c |} {0.31274}    \\
 \multicolumn{1}{| c }{} & \multicolumn{1}{l |}{1911 Schubart} & \multicolumn{1}{c |} {$ 3.14+  1.52\cos( 0.2964t+  3.01)$
 }  & \multicolumn{1}{c |} {0.32452}    \\
 \multicolumn{1}{| c }{} & \multicolumn{1}{l |}{448203 (2008UH167)} & \multicolumn{1}{c |} {$  3.05+  1.49\cos( 0.3180t+  4.37)$
 }  & \multicolumn{1}{c |} {0.34992}    \\
 \multicolumn{1}{| c }{} & \multicolumn{1}{l |}{46629 (1994 PS38)} & \multicolumn{1}{c |} {$  3.15+  1.49\cos( 0.3190t+  2.90)$
 }  & \multicolumn{1}{c |} {0.32684}    \\
 \multicolumn{1}{| c }{} & \multicolumn{1}{l |}{958 Asplinda } & \multicolumn{1}{c |} {$ 6.36+  1.60\cos( 0.3508t+  2.12)$
 }  & \multicolumn{1}{c |} { 0.29909}    \\
  \hline 
 \end{tabular}
 \end{center}
 \label{tabinc}
 \end{table*}

 \begin{table*}[t]
 \caption{The node longitude evolution approximation for the selected Hilda group members. Integration interval 100 kyr}
 \begin{center}
 \begin{tabular}{l l r r r } 
 \hline 
  \multicolumn{2}{| c |}{Asteroid }  &   \multicolumn{2}{c |}{}  \\
  \multicolumn{2}{| c |}{}&    \multicolumn{1}{c |}{$\Omega$}    &  \multicolumn{1}{c |} {$\sigma $}  \\
   \hline \hline 
 \multicolumn{1}{| c }{} & \multicolumn{1}{l |}{153 Hilda}   & \multicolumn{1}{c |}{$ 231.4-17.55t+   9.8\cos( 0.307t+  5.77)$}  & \multicolumn{1}{c |} {2.68476
 }   \\
 \multicolumn{1}{| c }{} & \multicolumn{1}{l |}{1180  Rita} & \multicolumn{1}{c |} {$  83.4-20.17t+  13.8\cos( 0.349t+  1.74)$
 }  & \multicolumn{1}{c |} {3.19927}   \\
 \multicolumn{1}{| c }{} & \multicolumn{1}{l |}{3990 Heimdal} & \multicolumn{1}{c |} {$ 201.4-18.77t+   8.9\cos( 0.326t+  5.98)$
 }  & \multicolumn{1}{c |} { 2.33797}  \\
 \multicolumn{1}{| c }{} & \multicolumn{1}{l |}{248430(2005 SM284)} & \multicolumn{1}{c |} {$68.9-15.70t+  8.9\cos( 0.270t+  2.10)$
 }  & \multicolumn{1}{c |} { 2.69237}    \\
 \multicolumn{1}{| c }{} & \multicolumn{1}{l |}{6124 Mecklenburg} & \multicolumn{1}{c |} {$ 167.9-16.50t+  10.2\cos( 0.289t+  0.65)$
 }  & \multicolumn{1}{c |} {2.76653}    \\
 \multicolumn{1}{| c }{} & \multicolumn{1}{l |}{1911 Schubart} & \multicolumn{1}{c |} {$ 290.0-17.10t+  29.4\cos( 0.301t+  4.90)$
 }  & \multicolumn{1}{c |} {8.99386}    \\
 \multicolumn{1}{| c }{} & \multicolumn{1}{l |}{448203 (2008UH167)} & \multicolumn{1}{c |} {$ 209.5-18.40t+  28.8\cos( 0.320t+  5.93)$}  & \multicolumn{1}{c |} { 8.67795}    \\
 \multicolumn{1}{| c }{} & \multicolumn{1}{l |}{46629 (1994 PS38)} & \multicolumn{1}{c |} {$ 299.2-18.22t+  29.3\cos( 0.319t+  4.53)$
 }  & \multicolumn{1}{c |} {8.76590}    \\
 \multicolumn{1}{| c }{} & \multicolumn{1}{l |}{958 Asplinda } & \multicolumn{1}{c |} {$ 330.0-20.34t+  14.0\cos( 0.352t+  3.88)$
 }  & \multicolumn{1}{c |} { 3.03681}    \\
  \hline 
 \end{tabular}
 \end{center}
 \label{tabNod}
 \end{table*}
 
 \subsection{The dependence frequencies on the type of perturbations}
 
{To study the dependence of our results on type of perturbations, we repeated our integration and approximation in case planets + Ceres + Vesta perturbations (Table \ref{tabExCV}). The effect of Ceres and Vesta perturbations on the inclination and node longitude evolution is negligible. But in case eccentricity and perihelion evolution this effect is small but detectable, up to $3.5\%$ in the long period value. In the most cases it leads to increasing the frequency of the short periodic perturbation and decreasing of the longperiodic frequency.}   

\subsection{The semimajor axis approximation}

{The semimajor axis is very sensitive to initial orbits variations and type of perturbations. by this reason we avoid to discuss it in our previous studying. But now we present result of our approximation of the semimajor axis of the selected members of Hilda group (Table \ref{tabAAx}). Note that only in case 153 Hilda the approximation by combined frequency by expression (\ref{Ei_mult}) is necessary.}

 \begin{table*}[t]
 \caption{The eccentricity evolution approximation for the selected Hilda group members}
 \begin{center}
 \begin{tabular}{l l r r r } 
 \hline 
  \multicolumn{2}{| c |}{Asteroid }  &   \multicolumn{2}{c |}{}  \\
  \multicolumn{2}{| c |}{}&    \multicolumn{1}{c |}{$e$}    &  \multicolumn{1}{c |} {$\sigma $}  \\
   \hline \hline 
 \multicolumn{1}{| c }{} & \multicolumn{1}{l |}{153 Hilda}   & \multicolumn{1}{c |}{$ 0.180+  0.074\cos( 2.460t+  1.90)\cos( 0.0615t+  5.68)$
 }  & \multicolumn{1}{c |} {0.01762}   \\
 \multicolumn{1}{| c }{} & \multicolumn{1}{l |}{1180  Rita} & \multicolumn{1}{c |} {$  0.173+  0.074\cos( 2.678t+  1.66)\cos( 0.0592t+  5.70)$
 }  & \multicolumn{1}{c |} {0.01765}   \\
 \multicolumn{1}{| c }{} & \multicolumn{1}{l |}{3990 Heimdal} & \multicolumn{1}{c |} {$ 0.170+  0.071\cos( 2.569t+  2.35)\cos( 0.0564t+  2.33)$
 }  & \multicolumn{1}{c |} { 0.01763}  \\
 \multicolumn{1}{| c }{} & \multicolumn{1}{l |}{248430(2005 SM284)} & \multicolumn{1}{c |} {$  0.203+  0.076\cos( 1.937t+  2.10)\cos( 0.0587t+  5.50)$
 }  & \multicolumn{1}{c |} {0.01802}    \\
 \multicolumn{1}{| c }{} & \multicolumn{1}{l |}{6124 Mecklenburg} & \multicolumn{1}{c |} {$  0.195+  0.077\cos( 2.018t+  2.09)\cos( 0.0587t+  2.44)$
 }  & \multicolumn{1}{c |} {0.01815}    \\
 \multicolumn{1}{| c }{} & \multicolumn{1}{l |}{1911 Schubart} & \multicolumn{1}{c |} {$  0.197+  0.073\cos( 2.140t+  4.04)\cos( 0.0588t+  5.50)$
 }  & \multicolumn{1}{c |} {0.01760}    \\
 \multicolumn{1}{| c }{} & \multicolumn{1}{l |}{448203 (2008UH167)} & \multicolumn{1}{c |} {$  0.185+  0.074\cos( 2.301t+  1.17)\cos( 0.0584t+  2.48)$
 }  & \multicolumn{1}{c |} {0.01760}    \\
 \multicolumn{1}{| c }{} & \multicolumn{1}{l |}{46629 (1994 PS38)} & \multicolumn{1}{c |} {$  0.185+  0.074\cos( 2.315t+  0.39)\cos( 0.0597t+  5.60)$
 }  & \multicolumn{1}{c |} {0.01750}    \\
 \multicolumn{1}{| c }{} & \multicolumn{1}{l |}{958 Asplinda } & \multicolumn{1}{c |} {$   0.178+  0.071\cos( 2.510t+  2.98)\cos( 0.0585t+  5.45)$
 }  & \multicolumn{1}{c |} {0.02215}    \\
  \hline 
 \end{tabular}
 \end{center}
 \label{tabEx}
 \end{table*}

 \begin{table*}[t]
 \caption{The perihelion longitude evolution approximation for the selected Hilda group members}
 \begin{center}
 \begin{tabular}{l l r r r } 
 \hline 
  \multicolumn{2}{| c |}{Asteroid }  &   \multicolumn{2}{c |}{}  \\
  \multicolumn{2}{| c |}{}&    \multicolumn{1}{c |}{$\omega$}    &  \multicolumn{1}{c |} {$\sigma $}  \\
   \hline \hline 
 \multicolumn{1}{| c }{} & \multicolumn{1}{l |}{153 Hilda}   & \multicolumn{1}{c |}{$ 247.3-136.31t+ 29.7\cos( 2.466t+  4.00)\cos( 0.0588t+  5.78)$
 }  & \multicolumn{1}{c |} {7.99984}   \\
 \multicolumn{1}{| c }{} & \multicolumn{1}{l |}{1180  Rita} & \multicolumn{1}{c |} {$ 271.3-148.71t+ 29.9\cos(2.676t+  3.20 )\cos(0.0584t+  5.60)$
 }  & \multicolumn{1}{c |} {7.91697}   \\
 \multicolumn{1}{| c }{} & \multicolumn{1}{l |}{3990 Heimdal} & \multicolumn{1}{c |} {$  45.1-142.59t+ 28.9\cos( 2.462t+  6.26)\cos( 0.0589t+  2.19)$
 }  & \multicolumn{1}{c |} {10.6983}  \\
 \multicolumn{1}{| c }{} & \multicolumn{1}{l |}{248430(2005 SM284)} & \multicolumn{1}{c |} {$ 223.2-106.37t+ 28.3\cos( 1.939t+  0.80)\cos( 0.0587t+  2.38)$
 }  & \multicolumn{1}{c |} {7.21058}    \\
 \multicolumn{1}{| c }{} & \multicolumn{1}{l |}{6124 Mecklenburg} & \multicolumn{1}{c |} {$  62.1-110.94t+ 27.7\cos( 2.019t+  3.67)\cos( 0.0586t+  2.45)$
 }  & \multicolumn{1}{c |} {7.34807}    \\
 \multicolumn{1}{| c }{} & \multicolumn{1}{l |}{1911 Schubart} & \multicolumn{1}{c |} {$ 141.4-117.64t+ 26.5\cos( 2.031t+  4.99)\cos( 0.0525t+  1.89)$
 }  & \multicolumn{1}{c |} {9.59673}    \\
 \multicolumn{1}{| c }{} & \multicolumn{1}{l |}{448203 (2008UH167)} & \multicolumn{1}{c |} {$ 118.0-127.19t+ 28.5\cos(2.286t+ 4.60 )\cos(0.0567t+  5.29)$
 }  & \multicolumn{1}{c |} {7.51839}    \\
 \multicolumn{1}{| c }{} & \multicolumn{1}{l |}{46629 (1994 PS38)} & \multicolumn{1}{c |} {$ 340.4-127.93t+ 28.3\cos( 2.310t+  1.47)\cos( 0.0572t+  5.41)$
 }  & \multicolumn{1}{c |} {7.30923}    \\
 \multicolumn{1}{| c }{} & \multicolumn{1}{l |}{958 Asplinda } & \multicolumn{1}{c |} {$  99.5-140.30t+ 27.0\cos( 2.510t+  1.40)\cos( 0.0553t+  2.20)$
 }  & \multicolumn{1}{c |} {9.47607}    \\
  \hline 
 \end{tabular}
 \end{center}
 \label{tabPeri}
 \end{table*}

   \begin{table*}[t]
    \caption{The eccentricity evolution approximation for the selected Hilda group members(Planets+Ceres+Vesta perturbations)}
    \begin{center}
    \begin{tabular}{l l r r r } 
    \hline 
     \multicolumn{2}{| c |}{Asteroid }  &   \multicolumn{2}{c |}{}  \\
     \multicolumn{2}{| c |}{}&    \multicolumn{1}{c |}{$e$}    &  \multicolumn{1}{c |} {$\sigma $}  \\
      \hline \hline 
    \multicolumn{1}{| c }{} & \multicolumn{1}{l |}{153 Hilda}   & \multicolumn{1}{c |}{$ 0.180+  0.074\cos( 2.460t+  1.96)\cos( 0.0609t+  5.68)$
    }  & \multicolumn{1}{c |} {0.01685}   \\
    \multicolumn{1}{| c }{} & \multicolumn{1}{l |}{1180  Rita} & \multicolumn{1}{c |} {$  0.173+  0.074\cos( 2.679t+  1.68)\cos( 0.0585t+  5.64)$
    }  & \multicolumn{1}{c |} {0.01683}   \\
    \multicolumn{1}{| c }{} & \multicolumn{1}{l |}{248430(2005 SM284)} & \multicolumn{1}{c |} {$  0.203+  0.076\cos( 1.937t+  2.09)\cos( 0.0586t+  5.40)$
    }  & \multicolumn{1}{c |} {0.01969}    \\
    \multicolumn{1}{| c }{} & \multicolumn{1}{l |}{6124 Mecklenburg} & \multicolumn{1}{c |} {$  0.195+  0.077\cos( 2.014t+  4.87)\cos( 0.0567t+  5.42)$
    }  & \multicolumn{1}{c |} {0.01850}    \\
    \multicolumn{1}{| c }{} & \multicolumn{1}{l |}{448203 (2008UH167)} & \multicolumn{1}{c |} {$  0.185+  0.074\cos( 2.303t+  1.18)\cos( 0.0578t+  2.47)$
    }  & \multicolumn{1}{c |} {0.01754}    \\
    \multicolumn{1}{| c }{} & \multicolumn{1}{l |}{46629 (1994 PS38)} & \multicolumn{1}{c |} {$  0.185+  0.074\cos( 2.317t+  0.43)\cos( 0.0586t+  5.50)$
    }  & \multicolumn{1}{c |} {0.01779}    \\
     \hline 
    \end{tabular}
    \end{center}
     \label{tabExCV}
    \end{table*}

   \begin{table*}[t]
    \caption{The semimajor axis evolution approximation for the selected Hilda group members}
    \begin{center}
    \begin{tabular}{l l r r r } 
    \hline 
     \multicolumn{2}{| c |}{Asteroid }  &   \multicolumn{2}{c |}{}  \\
     \multicolumn{2}{| c |}{}&    \multicolumn{1}{c |}{$a,AU$}    &  \multicolumn{1}{c |} {$\sigma $}  \\
      \hline \hline 
    \multicolumn{1}{| c }{} & \multicolumn{1}{l |}{153 Hilda}   & \multicolumn{1}{c |}{$ 3.965+  0.0015\cos( 2.32t+  3.00)\cos( 0.055t+  1.10)$
    }  & \multicolumn{1}{c |} {0.00067}   \\
    \multicolumn{1}{| c }{} & \multicolumn{1}{l |}{1180  Rita} & \multicolumn{1}{c |} {$  3.965+  0.0030\cos( 2.977t+  5.71)$
    }  & \multicolumn{1}{c |} {0.00071}   \\
    \multicolumn{1}{| c }{} & \multicolumn{1}{l |}{248430(2005 SM284)} & \multicolumn{1}{c |} {$  3.966+  0.0009\cos( 1.68t+  2.76)$
    }  & \multicolumn{1}{c |} {0.00050}    \\
    \multicolumn{1}{| c }{} & \multicolumn{1}{l |}{6124 Mecklenburg} & \multicolumn{1}{c |} {$  3.966+  0.0012\cos( 1.487t+  4.10)$
    }  & \multicolumn{1}{c |} {0.00056}    \\
    \multicolumn{1}{| c }{} & \multicolumn{1}{l |}{448203 (2008UH167)} & \multicolumn{1}{c |} {$  3.965+  0.0015\cos( 1.05t+  1.82)$
    }  & \multicolumn{1}{c |} {0.00060}    \\
    \multicolumn{1}{| c }{} & \multicolumn{1}{l |}{46629 (1994 PS38)} & \multicolumn{1}{c |} {$  3.965+  0.0015\cos( 0.92t+  4.40)$
    }  & \multicolumn{1}{c |} {0.00051}    \\
     \hline 
    \end{tabular}
    \end{center}
    \label{tabAAx}
    \end{table*}

\section{Discussion and conclusions}

The present paper provides some new quantitative data and draws attention to the difference in the dynamical state of the asteroids within the Hilda group. A brief overview of the obtained approximation allows us to make the following conclusions. 1) The eccentricities and perihelion longitudes of the studied asteroids are better approximated by combinational frequencies. This approximation determines the periods of perturbation more precisely than the Fourier method. 2) The Hilda and Schubart families are clearly separated not only by the mean inclination values but also by the amplitudes of the node longitude perturbation. 3) The main period in inclination and node longitude is the same for all studied asteroids. One of the periods of perturbations in the eccentricity and in the perihelion longitudes are the same for each asteroid. At the same time these periods are different for different members of Hilda group. 
	
Members of Hilda's group exhibit perturbations with slightly different frequencies and amplitudes. Understanding the cause of such variability may be the subject of future study.

We believe that our results can be useful in the study of the origin of the Hilda group, as well as for a better understanding of the details of the motion of small bodies in resonance.

\section{conflicts of interest}
The author declares that they have no conflicts of interest

\begin{thebibliography}{}
%
%
\bibitem[{Bailey and Fabrycky}(2020)]{BF}
Bailey N., Fabrycky D.,2020. \ Nodal Precession in Closely Spaced Planet Pairs. The Astronomical Journal, 159, 217 (13pp).
\bibitem[{Batygin, Morbidelli}(2013)]{BM}
Batygin K., Morbidelli A.,2013.\ Analytical treatment of planetary resonances. Astronmy and Astrophysics, 556, A28.
\bibitem[{Broz, Vokrouhlicky}(2008)]{BV}
Broz M., Vokrouhlicky D.,2008.\ Asteroid families in the first-order resonances with Jupiter. Mon. Not. R. Astron. Soc. 390, 715–732 
\bibitem[Celletti,et al.(2018)]{Ce1}
Celletti A., Paita F., Pucacco G.,2018.\ The dynamics of the De Sitter resonance., Celestial Mechanics and Dynamical Astronomy, 130, 2, article id. 15, 15 pp.
\bibitem[Chambers,(1999)]{Ch1}
Chambers, J. E.,1999.\ A hybrid symplectic integrator that permits close encounters between massive bodies. MNRAS, 304,793-799 
\bibitem[Everhart,(1985)]{E1}
Everhart, E.,1985.\ An efficient integrator that uses Gauss-Radau spacings. In A. Carusi, G. B. Valsecchi (Eds.), Dynamics of Comets: Their Origin and Evolution, Proceedings of IAU Colloq. 83, held in Rome, Italy, June 11-15, 1984. Edited by Andrea Carusi and Giovanni B. Valsecchi. Dordrecht: Reidel. Astrophysics and Space Science Library,115,185 
\bibitem[Knezevic, Milani(2003)]{KM}
Knezevic, Z., Milani, A.,2003. \ Proper element catalogs and asteroid families. Astronomy and Astrophysics, 403, 1165-1173.
\bibitem[Lainey et al(2006)]{La}
Lainey V., Duriez L., Vienne A., 2006.\ Synthetic representation of the Galilean satellites’ orbital motions from L1 ephemerides., Astronomy and Astrophysics, 456, 783–788.
\bibitem[Murray, Dermott,(1999)]{MD}
Murray, C. D., Dermott, S. F., 1999. \ Solar System Dynamics, Cambridge Univ. Press, Cambridge, 606 p..
\bibitem[Meletlidou, et al.,(2001)]{Me1}
Meletlidou E., Ichtiaroglou S., Winterber G F.J.,2001.\ Non–integrability of Hill’s lunar problem., Celestial Mechanics and Dynamical Astronomy., 80,   145–156.
\bibitem [Nesvorny (2018)]{NV2}
Nesvorny D., 2018.\ Dynamical Evolution of the Early Solar System. Annual Review of Astronomy and Astrophysics, vol. 56, p.137-174 
\bibitem[Rosaev (2022)]{R2}
Rosaev A.,2022.\ The resonance perturbations of the (39991) Iochroma family. Celestial Mechanics and Dynamical Astronomy, 134:48.
\bibitem[{Rosaev,Plavalova} (2021)]{RP5}
Rosaev A., Plavalova E., 2021.\ The Fourier approximation for orbital elements for the members of very young asteroid families, Planetary and Space Science, 202, 105233 
\bibitem[{Rosaev,Plavalova} (2022)]{RP}
Rosaev A., Plavalova E.,2022. \ Some of the most interesting cases of close asteroid pairs perturbed by resonance, Multi-Scale (Time and Mass) Dynamics of Space Objects. Held 18-22 October, 2021 in Iaşi, Romania. Proceedings of the International Astronomical Union, V. 364, pp. 226-231.
\end{thebibliography}

\end{document}